\documentclass[preprint,showpacs,preprintnumbers,amsmath,amssymb]{revtex4}
\usepackage{graphicx}% Include figure files
\usepackage{dcolumn}% Align table columns on decimal point
\usepackage{bm}% bold math

\newcommand{\be}{\begin{equation}}
\newcommand{\ee}{\end{equation}} \newcommand{\ba}{\begin{eqnarray}}
\newcommand{\ea}{\end{eqnarray}}
\begin{document}
\title{ Fermions in Deep Optical Lattice under p-wave Feshbach Resonance } \author{S.-K. Yip}
 \affiliation{ Institute of Physics, Academia Sinica, Nankang, Taipei 115, Taiwan} \date{\today}

\begin{abstract}
We investigate theoretically the energy levels of two identical
Fermions in a harmonic potential well under p-wave Feshbach
resonance.   The magnetic fields needed to affect the energy levels
differ strongly among the levels, and they can be at values far from
the free space resonant field.

 \end{abstract}

 \maketitle
%%%%%%%%%%%%%%%%%%%%%%%%%%%%%%%%%%%%%%%%%%%%%%%%%%%%%%%%%%%%%%%%
%This prediction can be tested experimentally
%by varying the magnetic field in time and observing the final
%occupation probabilities of the harmonic oscillator states.

There are substantial recent activities in studying cold atomic
gases in the presence of an optical lattice (e.g. \cite{Greiner02}).
Part of this interest stems from the possible connection with
solid-state physics (e.g. \cite{Hofstetter02}) and the vast
theoretical literature on exotic states in lattice hamiltonians.
Generally, the interaction between atoms are weak.  To increase the
interaction between particles, one possible way is to make use of
Feshbach resonance.  Indeed, a few such experiments have already
been carried out \cite{Kohl05,Stoferle06}.  An important question
arises already even when the tunneling of atoms between different
potential wells can be neglected.  For more than one particle in a
well, the energy levels, wavefunctions etc are strongly modified by
the Feshbach resonance. Theoretical studies of this "dressed state"
have been carried out in some recent papers
\cite{Dicker05,Diener06}. Experimentally, Ref \cite{Kohl05} has been
able to obtain some important information about this dressed state,
namely, the probability of finding a particle in a particular
harmonic oscillator level.  They did this by starting with a deep
lattice potential and, by sweeping the magnetic field through
resonance and then turning off the optical lattice, occupation of
the different Bloch bands was experimentally observed. The energy of
the lowest bound state has also been measured by radio frequency
spectroscopy \cite{Stoferle06}. Understanding these dressed state is
also crucial for obtaining the lattice Hamiltonian \cite{Duan05},
the first essential step to eventually able to understand any
many-body systems to be studied.

  The above references \cite{Kohl05,Stoferle06,Dicker05,Diener06,Duan05} are on
  s-wave Feshbach resonance between two Fermions of different
  internal hyperfine spins.  Here we would like to extend the study
  to two identical fermions (same hyperfine spin) under a p-wave
  Feshbach resonance.  P-wave Feshbach resonances in Fermions
  (e.g. \cite{Ticknor04,Zhang04,Gaebler07}) itself
  have generated much activities.  One reason is
  possible p-wave superfluidity
  (e.g. \cite{Ho-p,Gurarie05,Cheng05,Cheng06}).  There has also been a
  suggestion \cite{Jaksch02}  to obtain molecular superfluid by first starting with
  a lattice with two atoms per site and then "melt" the Mott
  state by turning off the optical lattice.  This may perhaps
  be a route to obtain p-wave superfluidity and gives additional
  impetus to study Fermionic atoms under p-wave Feshbach resonance
  within an optical lattice.

   In this paper, we shall then consider two identical Fermions in a
   harmonic potential well under a p-wave Feshbach resonance.  We
   shall find how the energy levels depend on the magnetic field
   detuning.  We shall show that, in contrast to s-wave resonances, due to the strong momentum
   dependence of the effective interaction, there is a large shift
   in the magnetic field from resonance which is necessary to affect the energy levels
   with the shift itself depending strongly on the energy level
   under study.  The validity of this prediction can be tested by
   performing an experiment similar to that of \cite{Kohl05}.
   We employed a method different from the existing literature, an approach
   some readers may find more transparent.
   We shall also report on the generalization to resonance with
   higher angular momenta.

   In a harmonic trap, we can first separate the center of mass
   versus the relative motion, and we need only to treat the latter.
   The wavefunction of the "dressed" molecule in the relative coordinate $\vec r$ consists of two parts,
   $\psi (\vec r)$ in the open channel and $\gamma \varphi_c(\vec r)$
   in the closed channel.  Here, for later convenience, we have
   introduced a normalized wavefunction $\varphi_c(\vec r)$ in the
   closed channel and denote the amplitude of our dressed molecule
   in this channel by $\gamma$.  The normalization of the full
   wavefunction is thus $|\gamma|^2 + \int_{\vec r} |\psi(\vec r)|^2
   = 1$, where $\int_{\vec r}$ is the short-hand for $\int d^3{\vec
   r}$.    For p-wave resonance, due to the dipole interaction, the
   resonant field depends on the angular momentum $m_l$ along the
   applied magnetic field direction.  For example, for $^{40}K$ in
   the $| \frac{9}{2}, -\frac{7}{2} > $ hyperfine state, the $m_l =
   0$ resonance occurs at a field around $0.5$G above the $m_l = \pm
   1$ ones \cite{Ticknor04}.  For definiteness, in below we shall assume that we have the $m_l = 0$
   resonance only.  Modifications need to other $m_l$'s are evident.

   The Schr\"odinger  equations are
   \begin{eqnarray}
   \left( - \frac{\hbar^2}{2 m_r} \nabla^2 + U (\vec r) \right) \psi
   (\vec r) + W (\vec r) \gamma \varphi_c(\vec r) &=& E \psi( \vec r)
   \label{S1} \\
   W(\vec r) \psi(\vec r) \qquad + \nu' \gamma \varphi_c(\vec r) &=& E \gamma
   \varphi_c(\vec r) \label{S2}
   \end{eqnarray}
    Here $m_r = m/2$ is the reduced mass, $U(\vec r)$ is the trap
    potential, and $\varphi_c(\vec r)$ is the closed-channel molecule of  $l = 1$, $m_l = 0 $
    symmetry with (unrenormalized) energy $\nu'$, and $W (\vec r)$
    is the coupling between the open and closed channels,
    and $E$ is the energy of the dressed state.
     Here, we
    have already truncated the Hilbert space for the closed-channel
    to one single state responsible for the resonance of interest.
    For simplicity, we have also suppressed the part of the
    wavefunction representing the internal (hyperfine) degrees of
    freedom which distinguishes between the open and closed
    channels.  Eq (\ref{S1}-\ref{S2}) are the same as those written
    down for the s-wave (e.g. \cite{Dicker05}) except the symmetry of
    $\varphi_c(\vec r)$.

    Equation (\ref{S1}) can be solved in terms of the Green's
    function defined by
    \begin{equation}
    \left( - \frac{\hbar^2}{2 m_r} \nabla^2 + U (\vec r) - E \right)
    G_E(\vec r, \vec r') =
   \frac{2 \pi}{m_r} \delta^3 (\vec r -  \vec r')  \ .
   \label{defG}
   \end{equation}
   Substituting the result into eq (\ref{S2}), we obtain the
   implicit equation for the energy $E$ as
   \begin{equation}
   E - \nu' + \frac{2 \pi}{m_r} \int_{\vec r, \vec r'}
     \tilde \varphi_c^* (\vec r) G_E (\vec r, \vec r') \tilde
     \varphi_c(\vec r') = 0
     \label{Eb}
  \end{equation}
  where we have introduced the short-hand $\tilde \varphi_c(\vec r)
  \equiv W (\vec r) \varphi_c(\vec r)$.  The problem thus reduces to
  finding $G_E(\vec r, \vec r')$. We here adopt a different method
    from ref \cite{Busch98}.  We would also like to express our final answer
    in terms of experimentally available quantities. For the latter,
    we first observe that, in the
    absence of the trapping potential, the expression in eq
    (\ref{Eb}) is related to a quantity that appears in the
    scattering amplitude.  Explicitly, the scattering amplitude for
    incoming (outgoing) wavevector $\vec k$ ($\vec k'$) can be found
    to be
    \begin{equation}
    f_{\vec k}(\vec k') = \frac{ - \frac{2 \pi}{m_r}
      \tilde \varphi_c^* (\vec k)  \tilde \varphi_c(\vec k')}
      { E - \nu' +  \frac{2 \pi}{m_r}  \int_{\vec r, \vec r'}
      \tilde \varphi_c^* (\vec r) G_E^{free} (\vec r, \vec r') \tilde
     \varphi_c(\vec r')}
     \label{fk}
     \end{equation}
     where $G_E^{free} (\vec r, \vec r')$ obeys eq (\ref{defG})
     with $U(\vec r)$ set to zero. Here
     $E \equiv \frac{\hbar^2 k^2}{ 2 m_r}$,
     $\tilde \varphi_c(\vec k) \equiv \int_{\vec r} \tilde \varphi_c (\vec
     r) e^{ - i \vec k \cdot \vec r}$.  For scattering in the $l=1$, $m_l = 0$ channel,
     $f_{\vec k}(\vec k')  = 4 \pi f_{10} (k) Y_1^{0 *} (\hat k) Y_1^0
     (\hat k')$ where $f_{10}$ has the following form at small
     energies:
     \begin{equation}
     f_{10}(k) = \frac{k^2}{ - \frac{1}{v} + c k^2 - i k^3}   \ .
     \label{f10}
     \end{equation}
     Here $v$ has the dimension of a volume and $c$ inverse length.
      These parameters, in principle available from experiment
      (e.g. \cite{Ticknor04})
        or other theoretical calculations, will be used as an input
        below.  Near the resonant field $B_0^*$, $- \frac{1}{v}$ is
        roughly linear in $(B-B_0^*)$, whereas $c$ can be regarded
        as  a constant.
     %We are using the same notation as in, e.g., \cite{Ticknor04}.

     At small energies, $\tilde \varphi_c (\vec k) \approx - i k_z
       \int_{\vec r} \tilde \varphi_c(\vec r) z $
        $\equiv - i k_z \sqrt{3} \alpha$, where we have
        also defined the coupling constant $\alpha$ (This definition
        is identical with that in \cite{Cheng05,Cheng06}).  We then
        have the correspondence
    \begin{equation}
    - \frac{m_r }{  2 \pi |\alpha|^2}
    \left[ E - \nu' + \frac{2 \pi}{m_r} \int_{\vec r, \vec r'}
     \tilde \varphi_c^* (\vec r) G_E^{free} (\vec r, \vec r') \tilde
     \varphi_c(\vec r') \right] =
     \left[ - \frac{1}{v} + c k^2 - i k^3 \right]
     \label{rel}
     \end{equation}
     This equation, when expanded and comparing powers of $k$, yields the same
     "renormalization" relations as those written before in
     momentum space in, e.g.,  \cite{Ho-p,Cheng05,Cheng06}.

     The Green's function $G_E (\vec r, \vec r')$ and
      $G_E^{free} (\vec r, \vec r')$ are both singular as $\vec r
      \to \vec r'$, diverging as $\frac{1}{ | \vec r - \vec r'|}$,
       but otherwise finite.
     Defining $G_E^{reg}(\vec r, \vec r') \equiv
      G_E(\vec r, \vec r') - G_E^{free} (\vec r, \vec r')$,
      $G_E^{reg}(\vec r, \vec r')$ is thus finite for all $\vec r$
      and $\vec r'$.  The equation (\ref{Eb}) for the energy can
      then be written as
      \begin{equation}
    - \frac{1}{v} + c k^2 - i k^3 = \frac{1}{|\alpha|^2}
      \int_{\vec r, \vec r'}
     \tilde \varphi_c^* (\vec r) G_E^{reg} (\vec r, \vec r') \tilde
     \varphi_c(\vec r')
     \label{Ereg}
     \end{equation}

      It can be verified that $G_E^{reg}$ obeys
      \begin{equation}
    \left( - \frac{\hbar^2}{2 m_r} \nabla^2 + U (\vec r) - E \right)
    G_E^{reg} (\vec r, \vec r') =
     - U (\vec r) G_E^{free} (\vec r, \vec r')  \ .
   \label{dGreg}
   \end{equation}
   Below, we shall find $G_E^{reg}(\vec r, \vec r')$ for the case of a spherical trap of
   frequency $\omega$, thus $U (\vec r) = \frac{1}{2} m_r \omega^2
   r^2$.
   It is sufficient to do this for
   $ r > r'$ since the $ r < r'$ part can be obtained by the symmetry
   of $G_E^{reg} (\vec r, \vec r')$.
   We have $G_E^{free}(\vec r, \vec r') = \frac{e^{i k |
   \vec r - \vec r'|}}{|\vec r - \vec r'|}$ $ = $
   $ 4 \pi i k \sum_{l,m_l} j_l (kr') h_l^{(1)} (kr) Y_{lm_l}(\hat r)
   Y_{lm_l}^*(\hat r')$ for $r > r'$.  By the symmetry of $ \tilde
   \varphi_c$, we see that only the terms with $l = 1$, $m_l=0$ would
   contribute.  Writing this part of $G_E^{reg} (\vec r, \vec r')$ as
   $g_1(r, r') Y_{10} (\hat r) j_1 (kr') Y_{10}^* (\hat r')$, we
   are thus left to solve
   \begin{equation}
   \left( - \frac{\hbar^2}{2 m_r} \frac{1}{r^2} \frac{ d}{dr} r^2
    \frac{d}{dr} + \frac{2}{ 2 m_r  r^2} + U (r) - E \right)
      g_1 (r, r') = - U (r) 4 \pi i k h_1^{(1)} (k r)
     \label{dg1}
     \end{equation}

   Fortunately, a particular solution for eq (\ref{dg1}) can easily
   seen to be $ - 4 \pi i k h_1^{(1)} (kr)$.  The origin of this is
   simply that $ - G_E^{free} (\vec r, \vec r')$ is a particular
   solution to eq (\ref{dGreg}).   The homogeneous solution to eq
   (\ref{dg1}) can be expressed in terms of the confluent
   hypergeometric functions $F$, so we obtain thus
   \begin{equation}
   g_1(r, r') = \left[ b_1(r') \frac{r}{l_r} F ( -\mu_1, \frac{5}{2},
      \left( \frac{r}{l_r} \right)^2 ) +
     b_2 (r') (\frac{l_r}{r})^2 F ( -\mu_2, -\frac{1}{2},
      \left( \frac{r}{l_r} \right)^2 ) \right]
       e^{ - \frac{1}{2}  \left( \frac{r}{l_r} \right)^2}
       - 4 \pi i k h_1^{(1)} (kr)
       \label{g1}
  \end{equation}
  where $\mu_1 \equiv \frac{E}{ 2 \omega} - \frac{5}{4}$,
  $\mu_2 \equiv \frac{E}{ 2 \omega} + \frac{1}{4}$
  $= \mu_1 + \frac{3}{2}$, and $l_r \equiv \sqrt{ \hbar/m_r \omega}$
  is
  the oscillator length for relative motion, and $b_1(r')$ and
  $b_2(r')$ are coefficients to be determined.  The condition that
  $G_E(\vec r, \vec r') \to 0$ as $r \to \infty$ gives
  $b_1 \frac{\Gamma(5/2)}{\Gamma(-\mu_1)} +
       b_2  \frac{\Gamma(-1/2)}{\Gamma(-\mu_2)} = 0$, that is
       $b_1 = \frac{8}{3} \frac{\Gamma(-\mu_1)}{\Gamma(-\mu_2)}
       b_2$.
  For small $r'$, $j_1(kr') \to k r'/3$.
  The condition that $G_E(\vec r, \vec r')$ must be regular in the
  limit $\lim_{r' \to 0} \lim_{ r \to  r'_{+}}$ therefore requires that
  there cannot be any $1/r^2$ term in $g_1 (r, r')$.  Using
  $h_1^{(1)} (kr) \to - i/ (kr)^2$ as $r \to 0$,
  we therefore have $ \lim_{r' \to 0} b_2 (r') =
    \frac{ 4 \pi}{ k l_r^2}$.   On the other hand, for small $r$, we
    obtain from eq (\ref{g1}) that $g_1(r, r') = b_1(r')
    \frac{r}{l_r} - \frac{ 4 \pi} {3} i k^2 r$. (It can be verified that, with the above
    value of $b_2$, there
    are no terms proportional to $r^{-1}$ or $r^0$ in $g_1(r,r')$).  Substituting this
    expression into the right-hand-side (RHS) of eq (\ref{Ereg}), we see that, in the
    limit of short-ranged $\tilde \varphi_c$, we need only the value of
    $b_1(r' \to 0)$, and the RHS can be evaluated to be
    $ \frac{ 3 k }{4 \pi} \frac{b_1(0)}{l_r} - i k^3$.
    The $- i k^3$ terms on the two sides of eq (\ref{Ereg}) cancel.
    Using the above obtained value for $b_1(0)$ and the definition for
    $\mu_1$ and $\mu_2$, we obtain finally the equation
   \begin{equation}
   - \frac{1}{v} + c k^2 = \frac{8}{l_r^3}
      \frac{\Gamma( \frac{5}{4} - \frac{E}{2 \omega})}
        {\Gamma( -\frac{1}{4} - \frac{E}{2 \omega})}
        \label{Efinal}
   \end{equation}
    for the energies of the dressed states, expressed entirely
    in terms of quantities entering the scattering amplitude
    (\ref{f10})
    and the harmonic trap frequency $\omega$ ($k^2 = 2 m_r E$).

\begin{figure}
%\special{psfile=env.eps voffset=-500 hoffset=40 hscale=60 vscale=60
%angle= -90} \vspace*{7.5in} \caption{caption} \label{fig1g0}
\includegraphics{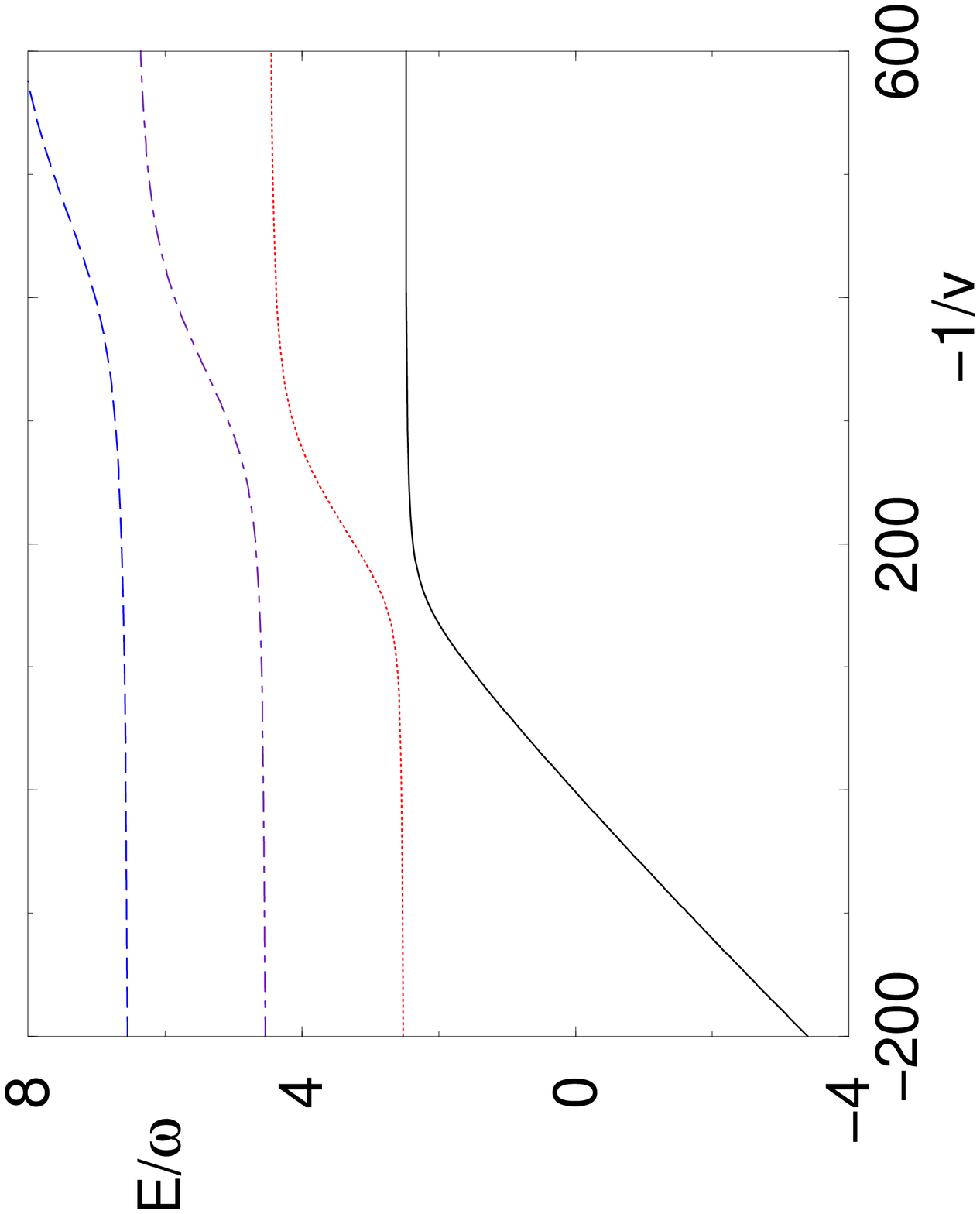}
\vspace*{6.5in} \caption{Energy levels versus ${-1/v}$.  Energies
are in units of the trap frequency, ${v}$ is unit of $l_{r}^3$.}
\label{fig:env}
\end{figure}

    It is convenient to rewrite this equation in dimensionless form:
   \begin{equation}
   - \frac{l_r^3}{v} - (- 2 c l_r)  \frac{E}{\omega}  = 8
      \frac{\Gamma( \frac{5}{4} - \frac{E}{2 \omega})}
        {\Gamma( -\frac{1}{4} - \frac{E}{2 \omega})}
        \label{Edimless}
   \end{equation}
   %The qualitative behavior of the solution can be seen graphically
   %by plotting both sides of this equation as a function of
   %$E/\omega$.
   For \cite{Ticknor04} $^{40}K$ $|\frac{9}{2},-\frac{7}{2}>$,
    $-c \approx 0.02 a_0^{-1}$, whereas $l_r \approx
   1600 a_0$ for a trap of frequency $ \omega = 70 {\rm kHz}$, a
   typical experimental number.  Thus the product
   $(- 2 c l_r)$ is large and $\approx 64$. The resulting energy
   levels are plotted in Fig \ref{fig:env}.  For "high" magnetic
   field,
   ($-\frac{1}{v} \to \infty$), the
   energy levels are given by the non-interacting values
   $(2n+\frac{5}{2}) \omega$ ($n$ is a non-negative integer).
    For $n = 0$, the state has
    energy decreasing with field, becomes linear in $-1/v$
    for $-1/v \lesssim 100$, and thus becomes indistinguishable with the
     the free space bound state energy \cite{Ho-p,Cheng05,Cheng06}
      $ - \epsilon_b =
    - \frac{\hbar^2}{2 m_r (-c v)}$ for $B < B_0^*$.
    For
   $n \ge 1$ and decreasing field, the energy of the
   level shift to $ (2 n + \frac{1}{2} ) \omega =
    (2 (n-1) +  \frac{5}{2} ) \omega$.    However, the value of the field
    where the energy crosses from its high field value to the low
    field one depends on the level $n$, increasing as $n$
    increases.  The width of field for this transition also
    increases with $n$. This is due to the large $ (- c l_r) $ value discussed
    above.

\begin{figure}
%\special{psfile=env.eps voffset=-500 hoffset=40 hscale=60 vscale=60
%angle= -90} \vspace*{7.5in} \caption{caption} \label{fig1g0}
\includegraphics{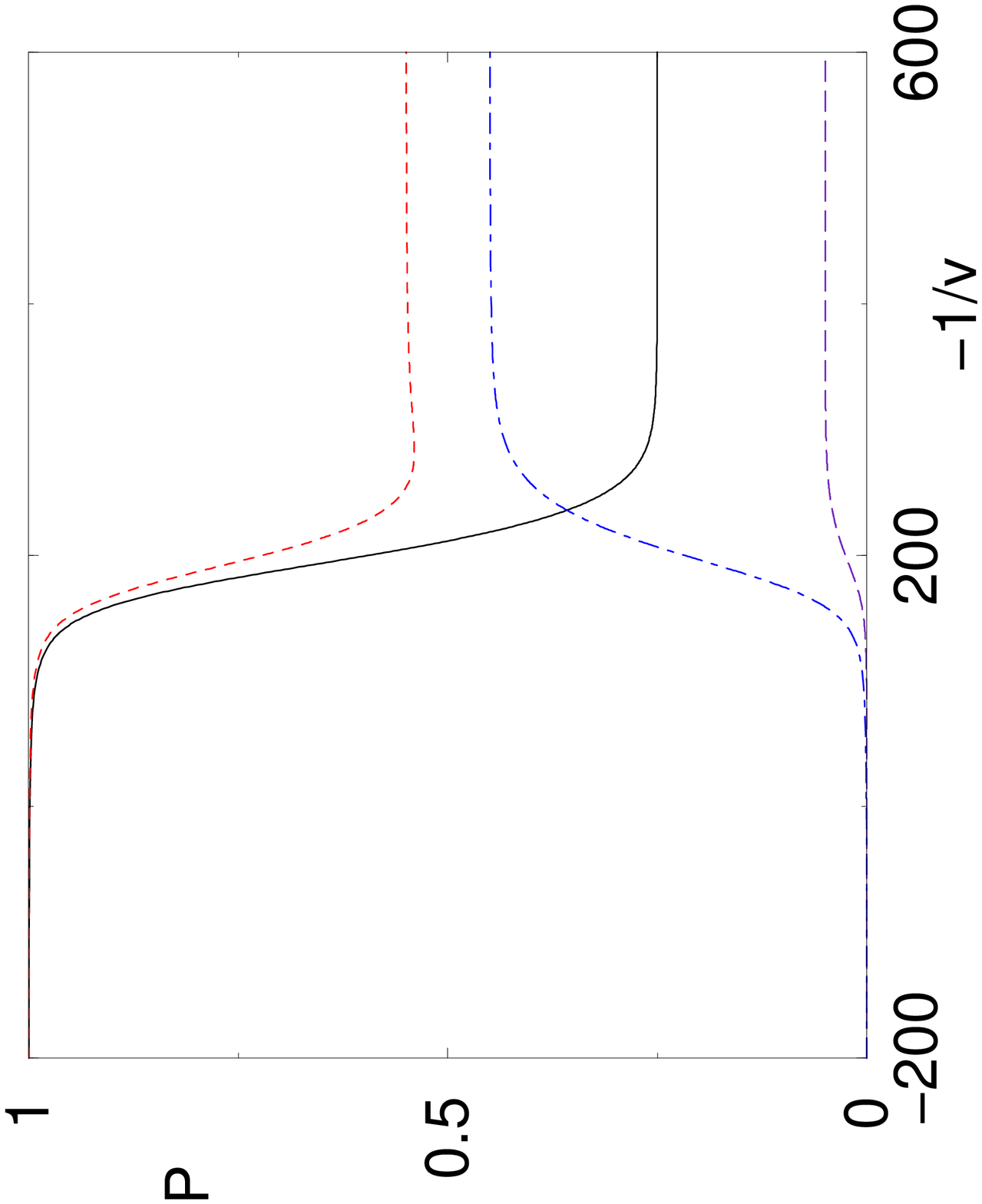}
\vspace*{6.5in} \caption{The probabilities $P$ of finding a particle
in the state $\vec n$ for $\vec n =$ $ (000)$ (solid), $(001)$
(dashed), $(100)$ (long dashed), and $(002)$ (dot-dashed), under the
field sweep procedure described in text.  ${v}$ is unit of
$l_{r}^3$.} \label{fig:P}
\end{figure}

     The energy levels can be measured by radio-frequency
     spectroscopy \cite{Stoferle06}.   Below, we analyze instead
     the implication on
     an experiment similar to that
    in \cite{Kohl05} (theoretically analyzed in \cite{Diener06,Katz06}),
     summarized already briefly  in the
    introduction.   We thus evaluate, for a given energy level of Fig \ref{fig:env},
     the probability of finding a
    particle in the single particle harmonic oscillator states.
    For this, we first have to express the two-particle wavefunction
    $\psi( \vec r)$ in harmonic oscillator basis.  A procedure
    similar to that in \cite{Diener06} gives $\psi (\vec r) = \sum_n
    \eta_n \psi _{n 1 0} (\vec r)$, where
    \begin{equation}
    \eta_n = \frac{ A_n / \left( \frac{E}{ 2 \omega} - ( 2 n +
    \frac{5}{2}) \right)}
  { \left\{ \frac{\pi^{3/2} l_r^5}{ 3 |\alpha|^2} +
  \sum_n \left[  A_n
   / \left( \frac{E}{ 2 \omega} - ( 2 n +
    \frac{5}{2}) \right) \right]^2 \right\}^{1/2} } \ ,
    \label{eta}
    \end{equation}
   $\psi_{n 1 0} (\vec r)
    \equiv  \frac{A_n}{\pi^{3/4} l_r^{5/2}} r {\rm cos} \theta
    F (-n, \frac{5}{2}, \left( \frac{r}{l_r} \right)^2)
      e^{ - \frac{1}{2} \left( \frac{r}{l_r} \right)^2 } $
    are the normalized wavefunctions with $l=1$, $m_l=0$,
    $A_n \equiv \left[ \frac{1}{3} \frac{ (2 n + 3) !!}{ 2^{n-1} n!} \right]^{1/2}$
    are normalization coefficients.  The amplitude in the closed
    channel is $\gamma = \left[ 1 - \sum_n |\eta_n|^2
    \right]^{1/2}$.  Below, we consider the special case where the
    term $\propto  \frac{1}{\alpha^2}$ in the denominator of eq
    (\ref {eta}) is negligible, correspondingly $\gamma \to 0$.
    (This assumption affects only quantitatively the probabilities
     at intermediate fields given below but not the qualitative conclusions.)
    Using this $\psi(\vec r)$ and assuming that the center of mass
    motion is in its harmonic oscillator ground state, it is
    straight-forward to express the resulting wavefunction in the
    coordinates $\vec r_1$, $\vec r_2$ of the two particles.
     For the latter wavefunction $\phi$, we shall employ instead the
     quantum numbers $n_x$, $n_y$ and $n_z$ since these are more
     directly related to the occupation of the different Bloch
     bands and thus the momentum states after turning off of the
     optical lattice.  For example, the state with quantum numbers $n_x$, $n_y$, $n_z$ contributes
     to momentum states with $  n_{x (y,z)} \frac{\pi}{d} <
     |k_{x (y,z)} | < (n_{x (y,z)}+ 1) \frac{\pi}{d}$, where $d$ is the
     distance between neighboring sites in the lattice \cite{Diener06}.
    The
    resulting expressions however are lengthy and we shall not write
    them down here, but only give the probabilities below.
    For definiteness, we consider an initial field $ B \ll B_0^*$
    with initial two-particle state corresponding to relative motion
    with quantum numbers $n = 0, l = 1, m=0$ (Energy of relative motion
    $=\frac{5}{2} \omega$.)  In the two-particle
    basis, the state is given by $ \frac{1}{\sqrt{2}}
    \left[ \phi_{000} (\vec r_1) \phi_{001} (\vec r_2) -
       \phi_{000} (\vec r_2) \phi_{001} (\vec r_1) \right]$
       where the subscripts stands for $(n_x, n_y, n_z) \equiv \vec n $.
   The initial probabilities of finding a particle in the single-particle state
   $\phi_{\vec n}$ are given by $P_{000} = 1$, $P_{001} = 1$.
   ( $P_{\vec n} = 0$ for all other $\vec n$).  The final probabilities of
   finding a particle in state $\vec n$ when the magnetic
   field is swept to the final value $B$ where the corresponding
   scattering parameter is
   $1/v$ is plotted for a few $\vec n$'s in Fig \ref{fig:P}.  The transition from the
   "low" field values to the "high" field values occur around
   $-l_r^3/v \approx 200$ ({\it c.f.} Fig \ref{fig:env}) for
   this particular state (This field increases with $n$).  For
   magnetic fields with $-1/v$ above this value, the probabilities
   correspond to the two-particle relative motion in the
   $n=1$, $l=1$, $m=0$ state.  We find $ P_{000} = \frac{1}{4}$,
   $ P_{001} = \frac{11}{20}$, $ P_{002} = \frac{9}{20}$,
    $ P_{100}= P_{010} = \frac{1}{20}$,
     $ P_{101}= P_{011} = \frac{1}{10}$, $ P_{003} = \frac{9}{20}$,
     with all other $P_{\vec n} = 0$.  The field at which the
     probabilities switch from one set of values to the other can
     be used to indicate when the energy levels change from the low
     to high field values.  For $^{40}$K in a trap with $70 {\rm kHz}$
     considered above, the shift $l_r^3/v \approx -200$ corresponds
     to $-\frac{1}{v} \approx 0.5 \times 10^{-7} a_0^{-3}$, or
     a field $\approx 1.15 {\rm G}$ above the free space resonance
     \cite{Ticknor04,footnote}, a large
     shift that should be readily discernible by experiments.
     This shift in field is roughly proportional to $1/l_r^2$ hence
     $\omega$.

     Finally, we discuss Feshbach resonances in higher angular
     momentum channels, say $l$, $m_l$ (Bosons or non-identical Fermions if $l$ even).
      Eq (\ref{fk}) is still valid,
     and the scattering amplitude
     has the form (generalizing eq (\ref{f10}))
     $k^{2l} / \left( - \frac{1}{a_{lm}^{2l+1}} + c_{lm} k^2 + ...
       - i k^{2l + 1} \right)$, where $a_{lm}$ has dimension of a length.
         One have relations similar to
       eq (\ref{rel}) and (\ref{Ereg}) with coupling constants
       $\alpha_{lm} = \frac{\sqrt{ 4 \pi}}{ (2 l + 1) !!}
       \int_{\vec r} \tilde \varphi_c (\vec r) r^l Y_{lm}^* (\hat r)$.
       We can obtain the corresponding part of $G_E^{ref}$ with
       relevant $l$, $m_l$ symmetry with
       $g_l(r, r')$ given by an expression similar to eq(\ref{g1})
       except that the $b_1(r')$ ($b_2(r')$) term has the factors
       $ \left(\frac{r}{l_r}\right)^l$
       ( $\left(\frac{r}{l_r}\right)^{-(l+1)}$)
        with appropriate changes in
       the arguments of the confluent hypergeometric functions $F$,
       and $h_1^{(1)} \to h_l^{(1)}$.
       If we proceed as in text and include only the regular term
       $\propto r^l$ in $g_l(r,r')$, we would get
       \begin{equation}
   - \frac{1}{a_{lm}^{2l+1}} + c_{lm} k^2  + ... =
     (-1)^{l+1} \left( \frac{2}{l_r} \right)^{2l+1}
      \frac{\Gamma( \frac{2l + 3}{4} - \frac{E}{2 \omega})}
        {\Gamma( -\frac{2l - 1}{4} - \frac{E}{2 \omega})}
        \label{El}
   \end{equation}
   This formula reduces to the known $l=0$ results in the literature
   \cite{Busch98,Dicker05,Diener06} and the $l=1$ result above (eq
   (\ref{Efinal})).  However, this is incomplete if $l \ge 2$.
   For $l = 2$ for example, the term $\propto b_2$, together with
   the $i n_2(kr)$ term from $h_2^{(1)}(kr)$, produce a term $\propto
    \frac{1}{r^3} \times r^4 = r$ as the lowest order contribution
    to $g_2(r,r')$.
    (It can be verified that there are no terms of lower order in
    $r$).  As a result, the RHS of eq (\ref{El}) acquires an extra
    term $R_{2m}$ given by $ - \frac{2 \Gamma (\frac{5}{2})}{l_r^4}
    \frac{ \int_{\vec r} \tilde \varphi_c (\vec r) r Y_{2m}^*(\hat
    r)}{\alpha_{2m}}$.  The integral involves short distance behavior
    of $\tilde \varphi_c$, and in contrast to the coupling constant $\alpha_{lm}$, cannot be
    eliminated from the final answer using the knowledge of the low energy scattering amplitude.
    Similar "residual" terms $R_{lm}$ arise for all $l \ge 2$, though the
    expressions become more involved and will not be given here.
    However, fortunately, for a given $lm$, $R_{lm}$ is independent of the
    energy, magnetic field (if near resonance) and the energy level
    under consideration. Since near the resonance, $\frac{1}{a_{lm}^{2l+1}}$ is
    expected to be linear in the deviation of $B$ from resonance, the extra
    term $R_{lm}$ simply gives a shift of
    the resonant field common to all the energy levels for given $lm$ and
    trap parameters $l_r$.  Apart from this,
    the behavior of the energy levels as a function of field is thus
    rather similar to that depicted in Fig \ref{fig:env}
    (though with energy levels given by $ (2 n + \frac{2 l + 3}{2}) \omega$ far from
    resonance).  The quantitative details will of course depend on
    the values of the parameters $c_{lm}$ etc.

    In conclusion,
    we have calculated energy levels of two Fermions in a harmonic potential well
    under a finite angular momentum Feshbach
      resonance. In contrast to wide s-wave Feshbach resonances,
      there are large shifts in the resonant fields dependent on the
      energy levels.  This prediction can be tested by experiments
      similar to those already performed in
      \cite{Kohl05,Stoferle06}.

       This research is supported by the National Science Council of
       Taiwan under grant number NSC 95-2112-M-001-054-MY3.

%%%%%%%%%%%%%%%%%%%%%%%%%%%%%%%%%%%%%%%%%%%%%%%%%%%%%%%%%%%%%%%%%%%%

%%%%%%%%%

%%%%%%%%%%%%%%%%%%%%%%%%%%%%%%%%%%%%%%%%%%%%%%%%%%%%%

%%%%%%%%%%%%%%%%%%%%%%%%%%%%%%%%%%%%%%%%%%%%%%%%%%%%%

\newpage

%%%%%%%%%%%%%%%%%%%%%%%%%%%%%%%%%%%%%%%%%%%%%%%%%%%%%%%%%%%

\newpage

\begin{thebibliography}{99}
\bibitem{footnote} See also Fig 2 in ref \cite{Cheng06}.  The unit
of the y-axis there should read $10^{-7} a_0^{-3}$. The correct
values were however used in the rest of that paper.
\end{thebibliography}
\end{document}